\documentstyle[12pt]{article}
\newcommand{\psib}{\bar{\psi}}
\newcommand{\Psib}{\bar{\Psi}}
\newcommand{\partbar}{\partial \!\!\! / \;}
\newcommand{\sigmab}{\bar{\sigma}}

\begin{document}
\title{Continuous media interpretation of  supersymmetric Wess-Zumino
 type models}
\author{P.S. Letelier\thanks{e-mail:letelier@ime.unicamp.br}
\\
Departamento de Matem\'atica Aplicada-IMECC\\
Universidade Estadual de Campinas\\
13081 Campinas. S.P., Brazil
\and
V.T. Zanchin\thanks{e-mail:zanchin@super.ufsm.br}
\\
 Departamento de F\'{\i}sica-CCNE
 \\Universidade Federal de Santa Maria \\
 97119 Santa Maria, R.S., Brazil\\}
\date{}
\maketitle
\begin{abstract}
\baselineskip 0.6cm

Supersymmetric  Wess-Zumino type models are considered as
classical material media that can be interpreted  as  fluids
of ordered strings with heat
flow along the strings or a mixture
of fluids of ordered strings with either a cloud of particles
or a flux of directed radiation.
\end{abstract}
\newpage
\baselineskip 0.8cm
{\it Introduction}.
The continuous media interpretation of a given field theory has been of
great utility in finding specific models of either perfect or anisotropic
fluids with simple physical and mathematical  properties. For instance,
it is well known that scalar fields may be used to model different material
media such as irrotational perfect fluids \cite{taub} and  anisotropic
fluids \cite{let1}.
Non-linear sigma models may model topological defects as cosmic strings
\cite{gibbons}, cosmic walls  and bubbles \cite{letverd}.

Following the same idea, Dirac and Weyl spinor fields also have been used
to construct models of continuous matter. In particular, massive neutrinos
can   behave  as perfect fluids \cite{wills}. In general, these fields
 can be
used to represent non-perfect fluids  \cite{cm,dr,spinors}.
Furthermore, anisotropic fluids with interesting physical properties
 and
fluids of cosmic strings with heat flow along the strings can also be
 constructed
from non-compact sigma models \cite{plvz}.

In such a formalism the fields are considered as potentials that describe the
fluids, i.e., the usual quantum field theory interpretation is abandon. Then,
following
this principle, it is a natural question to ask about the  ``fluid
interpretation" of supersymmetric fields. In this paper we study the
 Wess-Zumino model
 and following the principle of
considering each field as a matter potential we construct
some physically meaningful material media.  We would like to mention  that
a model of supersymmetric fluid that does not follow the underlying
principle mentioned above has already been discussed \cite{kuper}. This
 model,
as the author mention,  has not  a clear physical interpretation.

 We show that it is possible to have an interpretation of
the supersymmetric Wess-Zumino model as continuous media if the fields
are taken classically. In such a case, the spinor field does not
contribute to the metric (symmetrized) energy-momentum tensor (EMT) being
a {\it gravitational  ghost field}\, \cite{dr}. The resulting EMT is due only
to the scalar and
pseudo-scalar fields of the model. Then, in order to have a contribution
to the EMT from the spinor field,  it is necessary to introduce
in the standard Wess-Zumino Lagrangean  terms that are usually
neglected, as total divergences or  non-renormalizable (even though
regularizable) terms \cite{weinb}. It is shown that when such terms are
taken into account the spinor field can   have a significant contribution
 in the resulting EMT.

  Considering a particular interaction term we find that it is possible
to construct models of anisotropic fluids wherein the pressure along the
anisotropy direction (perpendicular) is smaller than  the
isotropic (parallel) pressure. This fact is of particular interest because it
represents a physical condition necessary to model objects  in astrophysics
\cite{hsw} and  cosmology \cite{goetz}.
 A second  model that can be constructed from the present approach
is a cosmic string fluid with heat flow along the strings. Strings with heat
flow along its length have been considered in the context of geometric
extensions of Nambu strings \cite{bl}.
It is also found models that are a mixture of cosmic string fluid with either
a cloud of particles or a flux of directed radiation.

{\it The energy-momentum tensor of Wess-Zumino model}. The supersymmetric
Wess-Zumino model is given by the following Lagrangean
\cite{wz}
\begin{equation}
L = {1\over {2}}[A^{,\mu}A_{,\mu} + B^{,\mu}B_{,\mu}+\Psib(i\partbar -
        m)\Psi -g\Psib(A+i\gamma^{\bf 5}B)\Psi- V(A,B)], \label{lgr}
\end{equation}
with
\begin{equation}
V(A,B) \equiv m^{2}(A^{2}+B^{2})+2mgA(A^{2}+B^{2})+g^{2}(A^{2}+
B^{2})^{2}\, ,  \label{pot}
\end{equation}
where $A$ and $B$ are both Hermitians spin-0 fields (scalar and
pseudo-scalar, respectively) and $\Psi$ is a spin-1/2 Majorana field.
 The base space of the model is the superspace \cite{sal},
being the sector related to the space-time coordinates the usual Minkowski
space.  The  results that we shall be  concern in this paper do not depend
on this particular choice
and may be easily generalized in the usual way to a local
Lorentzian subspace what makes possible the coupling with Einstein
equations.

The Lagrangean (\ref{lgr}) in fact corresponds to the on shell
 representation of
the model where the fields satisfy the equations
\begin{eqnarray}
& &( \Box + m^{2})A = -g\left[\Psib\Psi+m\,A(3\, A^{2}+B^{2})
    +2\,g\,A(A^{2}+B^{2})\right]\, , \\
& &( \Box+m^{2})B = -g\left[i\Psib\gamma^{5}\Psi+2\,m\,A\,B
               +2\,g\,B(A^{2}+B^{2})\right] \, \\
& &(i\partbar - m)\Psi = 2\,g(A +i\gamma^{5}B)\Psi\, .
\end{eqnarray}

We may consider in
the Lagrangean (\ref{lgr}) additional total divergence terms
(topological terms) as
\begin{equation}
L_{D} = {\Omega^{\mu}}_{,\mu}\; ,  \label{tdiv}
\end{equation}
that without changing the dynamics of the model can  yield non-trivial
changes in the EMT. For instance, (\ref{tdiv}) contributes with
 a new term of
the form \cite{sedov}
\begin{equation}
T_{\mu\nu} = \Omega_{\mu,\nu}+\Omega_{\nu,\mu},
\end{equation}
which may  modify the resulting continuous media
 changing for instance the state equations.

Another  way to modify the resulting EMT without breaking the
supersymmetric character of the model is by considering in the Lagrangean
non-renormalizable terms. Following Weinberg \cite{weinb} we may also
consider ``regularizable" terms which do not break the symmetry of
the model and there is no physical reason to  neglect. Such terms
can be written using the ``invariants" constructed with the conserved
currents of the original model.

For our purposes it is also interesting to consider other kinds
of non-renormalizable term like
\begin{equation}
L_{I} = \alpha\, A_{,\mu}\Psib\gamma^{\mu}\Psi\, .\label{lgri}
\end{equation}
This simple term is particularly interesting because when $g=0$ and $m=0$
[cf. (\ref{lgr})] is an invariant constructed with two conserved currents:
$I^{\mu} = A^{,\mu}$ and $J^{\mu}= \Psib\gamma^{\mu}\Psi$.

 Let us  write the (symmetrized) EMT obtained from (\ref{lgr}) with the
additional term $L^{1}$
\begin{eqnarray}
T_{\mu\nu} = A_{,\mu}A_{,\nu} + B_{,\mu}B_{,\nu}-
 \frac{1}{2} g_{\mu\nu}[A^{,\rho}A_{,\rho}
    + B^{,\rho}B_{,\rho} - V(A,B)] \nonumber \\ +\frac{
i}{8}(\Psib\gamma_{\mu}
    \Psi_{,\nu}-\Psib_{,\nu}\gamma_{\mu}\Psi +\Psib\gamma_{\nu}\Psi_{,\mu}
 -\Psib_{,\mu}\gamma_{\nu}\Psi) +T^{1}_{\mu \nu}\, .\label{emt0}
\end{eqnarray}
where $T^{1}_{\mu\nu}$ stands for the part of the EMT associated with
 (\ref{tdiv}) and/or (\ref{lgri}).

To have a consistent  supersymmetric theory   the components of the spinor
field $\Psi$ must be anticommuting Grassmann numbers. This  implies
that the EMT (\ref{emt0}) cannot be considered as a source of the classical
gravitational field, because its components belong to the Grassmann algebra
while the components of usual  Einstein equations are real.
In order to overcome this difficulty, it is  usually taken the {\it expected
values} of the energy-momentum tensor operator $T_{\mu\nu}$ as the source of
the Einstein equations. In this paper we are interested in the
interpretation of each field as a ``matter potential", in the sense of
the Debye potentials for fluids. In other words, we shall rewrite (\ref{emt0})
in an appropriated form and compare it to the energy-momentum tensor of a
``fluid". A simple way of solving this problem for the present case
is to consider all the fields of the model
classically, i.e., the fields $A$ and $B$ as scalar fields,
while the spinor field $\Psi$ as a classical (commuting) Majorana field with
 c-number components. We shall adopt this strategy and call the
resulting model as the ``classical interpretation of the Wess-Zumino model".

An important property of the classical Majorana spinor field is that the
corresponding symmetrized EMT vanish identically, i.e., it is a
{\it gravitational ghost field} \cite{cm,dr}. In the Weyl
representation the commuting bi-spinors satisfy \cite{mkw},
$\psib\sigmab_{\mu}\psi_{,\nu} = \psi_{,\nu}\sigma_{\mu}\psib$. Then
identical terms in (\ref{emt0}) cancel and the
resulting EMT for the classical interpretation of the {\it on shell}
Wess-Zumino model reduces to
\begin{equation}
T_{\mu\nu} = A_{,\mu}A_{,\nu}+B_{,\mu}B_{,\mu}  -\frac{1}{2} g_{\mu\nu}
\left[A^{,\rho}A_{,\rho} + B^{,\rho}B_{,\rho} -V(A,B)\right]
+T^{1}_{\mu \nu}\, ,  \label{emt} \end{equation}

The contribution of the Majorana spinor field appears  in
$T^{1}_{\mu\nu}$, i.e., in the case that $T^{1}_{\mu\nu} = 0$, the
spinorial part
of the usual supersymmetric Wess-Zumino model does not contributes to the
resulting continuous media that follows from (\ref{emt}). However,
contribution of this field can  appear when we consider interactions or
topological terms as the above mentioned. To be more specific
we choose the particular  Wess-Zumino model (\ref{lgr}) with
the interacting term (\ref{lgri})]. In such a case Eq. (\ref{emt}) reads
\begin{equation}
T_{\mu\nu}=A_{,\mu}A_{,\nu}+B_{,\mu}B_{,\nu} -\frac{1}{2}
 g_{\mu\nu}[A_{,\rho}A^{,\rho}
 + B_{,\rho}B^{,\rho}-V(A,B)] +\frac{1}{2}\alpha[A_{,\mu}J_{\nu}+
 A_{,\nu}J_{\mu}]\, ,
 \label{tmnf}
\end{equation}
 where $\alpha$ and $J_{\mu}$ are the same already defined quantities.\\

{\it Matter interpretation of the model}. A particularly
simple, albeit interesting, example is obtained
assuming  $A_{,\nu}A^{,\nu} > 0$ and  $B_{,\nu}B^{,\nu} > 0$ in (\ref{tmnf}).
In this case we can  define two timelike unit vectors by
\begin{equation} \begin{array}{cc}
  u_{\mu} = A_{,\mu}/\sqrt{A_{,\nu}A^{,\nu}}\, , &
  v_{\mu} = B_{,\mu}/\sqrt{B_{,\nu}B^{,\nu}}\, . \end{array}
  \label{uv}   \end{equation}
Moreover, the lightlike  or null vector $J_{\mu}$ $(\equiv\Psib\gamma_{\mu}
\Psi)$ may be decomposed as
\begin{equation}    \begin{array}{cc}
J_{\mu}= N(u_{\mu}+z_{\mu})\, ,
 & N  = J_{\mu}u^{\mu} = J_{\mu}A^{\mu}/\sqrt{A_{,\nu}A^{,\nu}}\, ;
    \end{array} \label{ax}
\end{equation}
where the spacelike vector $z_{\mu}$ obeys,
\begin{equation} \begin{array}{cc}
u^{\mu}z_{\mu}= 0\, , &  z^{\mu}z_{\mu}= -1\, \end{array} \label{x}.
\end{equation}
Therefore  the EMT (\ref{tmnf}) can be written into the form
\begin{equation}
T_{\mu\nu}=(r+2q)u_{\mu}u_{\nu}+sv_{\mu}v_{\nu}+q[u_{\mu}z_{\nu}+
  u_{\nu}z_{\mu}] -\frac{1}{2} g_{\mu\nu}[r +s -V(A,B)]\, ,    \label{tmnl1}
\end{equation}
where we have introduced the definitions
\begin{equation}   \begin{array}{ccc}
r \equiv A_{,\nu}A^{,\nu}\, ,  & s \equiv B_{,\nu}B^{,\nu}\, ,
 & q \equiv\frac{1}{2} \alpha A_{,\mu}J^{\mu}\, . \end{array}
    \label{qrs}    \end{equation}
Now, the EMT (\ref{tmnl1}) may be easily cast into its canonical
forms \cite{he} and interpreted as describing different continuous media
depending mainly upon the constant
$\alpha$. The algebraic structure of such a tensor can be analyzed
via its  associated secular or eigenvalue equation which in the
present case is equivalent the quartic equation\\
 $$\Lambda \{ \Lambda^{3}-\Lambda^{2}(r+s+2q)+
\Lambda[q^2+(r+2q) s ( 1-(u_{\mu}v^{\mu})^{2})-
2qs(v_{\mu}z^{\mu})(v_{\nu}u^{\nu})]  $$
\begin{equation}
+q^{2}s((v{_\mu}u^{\mu})^{2}-
(v_{\mu}z^{\mu})^2-1) \}=0  \, ,\label{eigeq}
\end{equation}
where the eigenvalues  $\lambda$  of (\ref{tmnl1})   are related to $\Lambda$
by $\lambda = \Lambda -(r+s-V)/2$. From Eq. (\ref{eigeq}) we have
the root $\Lambda=0$,
i.e., $2\lambda_{1} = - r-s+V$. The explicit expressions for the other three
eigenvectors are not  simple and they shall be presented elsewhere. In this
paper we shall rather study some significant particular  cases. \\

{\bf (i)} From (\ref{eigeq}) we see that  when  $q = 0$ and  $A_{,\mu}
\neq 0$ the resultant EMT represents the sum of two irrotational perfect
fluids. This particular case  was already studied by one of us
in  a different  context  \cite{let1}.

{\bf (ii)} Another simple situation is  when both vectors $v_{\mu}$ and
$u_{\mu}$ are parallel each other, i.e., when $u^{\mu}v_{\mu} =1$ and
consequently $z^{\mu}v_{\mu}=0$. In this case the resultant EMT is
equivalent to the one  of a nonviscous fluid with heat flow which can be of
interest in some applications, for instance, to construct models of
anisotropic universes and anisotropic fluid spheres \cite{hsw}, and also to
describe the non-adiabatic collapse of relativistic stars \cite{nos}. Under
the mentioned conditions the eigenvalue equation reduces to
\begin{equation}
{\Lambda}^{2} \left[\Lambda^{2} -\Lambda(r+s+2q)+q^{2}\right]=0\, .
            \label{eigeqa}  \end{equation}
This last equation yields four eigenvalues, two of them equal,
$\lambda_{1}=\lambda_{2}=-(r+s-V)/2$, and
[$\lambda_{\pm}=\Lambda_{\pm}-(r+s-V)/2$] given by
\begin{equation}
\lambda_{\pm}= (2q+V \pm \sqrt{(r+s+2q)^{2}-4q^{2}})/2 \, ,
    \label{eigenval}  \end{equation}
Then, it follows that $\Lambda_{\pm}$ may be either real if $(r+s+2q)^{2}
\geq 4q^{2}$ or complex if $(r+s+2q)^{2} < 4q^{2}$. In other words we have
three different cases.\\

{\em a)} First we consider the case with $\Lambda_{+} \geq \Lambda_{-}$.  If
we further
assume that $V = V(A,B) \geq 0$,   the relation
$\lambda_{+} \geq\lambda_{-}$ holds and
  $\lambda_{+} $ is associated to the timelike  eigenvector of
(\ref{tmnf}), and $\lambda_{-}$ to a spacelike eigenvector.
In this  case the EMT (\ref{tmnl1}) can be cast in the  canonical form
\begin{equation}
   T_{\mu\nu}=(\rho+\pi)U_{\mu}U_{\nu} +(\sigma-\pi)X_{\mu}X_{\nu}
     -\pi g_{\mu\nu}\, , \label{emtanis}
\end{equation}
where
\begin{eqnarray}
\rho &=&[2q+V+\sqrt{(r+s+2q)^{2}-4q^{2}}]/2 \, , \nonumber \\
\sigma &=&[-2q-V + \sqrt{(r+s+2q)^{2}-4q^{2}}]/2\, , \nonumber \\
\pi &=& (r+s-V)/2\, , \nonumber \\
U_{\mu} &=&C_{+}u_{\mu}+D_{+}v_{\mu}\, , \nonumber \\
X_{\mu} &=&C_{-}u_{\mu}+D_{-}v_{\mu}\,. \label{anisvec}
\end{eqnarray}
With
\begin{equation} \begin{array}{lr}
   C_{\pm} \equiv \Lambda_{\pm}\sqrt{|{\Lambda_{\pm}}^{2}-q^{2}|\,}, \,\,\,\,
&
   D_{\pm} \equiv -q\sqrt{|{\Lambda_{\pm}}^{2}-q^{2}|\,}, \end{array}
 \label{coeff}
\end{equation}
and  $\Lambda_{\pm} = \lambda_{\pm} +\pi$. One can verify that the
eigenvectors
$U_{\mu}$ and$X_{\mu}$ are orthonormal, i.e.
\begin{equation}
U_{\nu}U^{\nu}=-X_{\nu}X^{\nu} =1; \hspace{2cm} X_{\nu}U^{\nu} =0\, .
  \label{onorm}
  \end{equation}
  The EMT (\ref{emtanis}) is the
energy-momentum tensor of an anisotropic fluid with pressure
$\sigma$ along $X_{\mu}$ (the direction of the anisotropy) and pressure $\pi$
on the perpendicular plane to the anisotropy.

The variable $\sigma$  can also take   negative values. In
such a situation ( tension along the anisotropy direction) Eq. (\ref{emtanis})
may be put  in the form
\begin{equation}
   T_{\mu\nu}=(\lambda+\pi)(U_{\mu}U_{\nu}- X_{\mu}X_{\nu})- \pi g_{\mu\nu}\,
   + \varrho\, U_{\mu}U_{\nu}, \label{emtstr1}
\end{equation}
where $\lambda = -\sigma$ and $\varrho=\rho+\sigma=\sqrt{(r+s+2q)^2-4q^2}$\, .
The first part of the right hand side of Eq. (\ref{emtstr1}) is a fluid of
ordered strings \cite{matlet} while the second  is a cloud of particles.
Fluids of the kind represented by (\ref{emtstr1}) are a simple
generalizations of the clouds of ``realistics" strings considered in
 \cite{letsc}.\\

{\em b)} In the case $(r+s+2q)^{2} \leq 4q^{2}$, the tensor (\ref{tmnl1})
has  the  canonical form
\begin{equation}
T_{\mu\nu}=(\rho+ \pi)(U_{\mu}U_{\nu}-X_{\mu}X_{\nu})  -\pi g_{\mu\nu}
 + \beta (U_{\mu}X_{\nu}+U_{\mu}X_{\nu})  \, ,\label{emtstr}
\end{equation}
where
\begin{eqnarray}
\rho &=& (2\, q + V)/2\,, \nonumber\\
\pi & = &(r+s-V)/2 \, , \nonumber \\
\beta &=& \frac{1}{2}\sqrt{4\,q^2 -(r+s+2\,q)^2}\, ,\label{strvec}
\end{eqnarray}

The  EMT (\ref{emtstr}) represents  a   fluid made of parallel strings
with a heat flow $q_\mu=\beta X_\mu$ along the strings. $U_{\mu}$ is a
timelike vector while $X_{\mu}$ is spacelike and satisfies the orthonormality
conditions (\ref{onorm}). Their relations with the initial fluid
quantities are quite involved and they will be presented elsewhere.
It is worth to mention that strings with  heat flow along its length arise
naturally
when one considers  cosmic strings described by the Nambu action corrected by
terms built with the curvature of the string world sheet, these corrections
represent a geometric description  of the string width \cite{bl}.

{\em c)} In the case $(r+s+2q)^{2} = 4q^{2}$ the
canonical form of (\ref{tmnl1}) is
\begin{equation}
T_{\mu\nu}=(\rho+ \pi)(U_{\mu}U_{\nu}-X_{\mu}X_{\nu})  -\pi g_{\mu\nu}
 + \beta J_{\mu}J_{\nu}  \, ,\label{emtstrn}
\end{equation}

The EMT (\ref{emtstrn}) is the sum of a fluid of ordered strings and a null
fluid, i.e., a fluid of radiation directed along the strings. This situation
may be interesting to generalize the model of interacting strings and
electromagnetic radiation considered in \cite{let2}.\\

Similar results to ones presented in
the cases {\em a}, {\em b} and {\em c} above
have been also found considering the matter interpretation of non-compact
sigma models \cite{plvz}. Besides the cases above considered
there are other eight different situations, namely, the quantities
$A_{,\mu}A^{\mu}$ and $B_{,\mu}B^{\mu}$ may be positive, negative or zero.
{}From this we must consider all possible (different) combinations of these
values, which yields nine cases that should be studied separately.
The algebraic structure of the EMT (\ref{tmnf}) is different for each case,
and so are the resultant (if any) fluid interpretations. However, we
restricted ourselves to the case presented above because it
seems to be the ones with more physical content.

{\it Conclusions}.
We have shown that the supersymmetric Wess-Zumino model may be used as
motivation to find  models of continuous matter.

New simple continuous matter models can be obtained when total
divergences or other
non-renor\-malizable interaction terms are considered. In particular, the
presence of the interaction term (\ref{lgri}) made possible to find
different models:
{\bf i)} The anisotropic fluid (\ref{emtanis}) with anisotropic pressure
$\sigma$ (along the direction of anisotropy ($X_{\mu}$) smaller than the
isotropic pressure $\pi$. $\sigma$ may even be negative, representing a
tension along the anisotropy.
{\bf ii)} A fluid of ordered strings with heat flow along the strings, cf.
Eq. (\ref{emtstr}).
{\bf iii)} A mixture of a fluid of ordered strings with a fluid of directed
 radiation along the strings. This model follows from  Eq. (\ref{tmnl1}) in
the degenerated case where $(r+s+2q)^{2} = 4q^{2}$.

\end{document}